\documentclass[aps,prl,twocolumn,amsmath,nofootinbib]{revtex4-1}
\usepackage{bm}
\usepackage[toc,page]{appendix}
\usepackage{amsmath} 
\usepackage[margin=2.5cm]{geometry} 
\usepackage{mathtools}
\usepackage{caption}
\usepackage{subcaption}
\usepackage{array,multirow}
\usepackage[usenames, dvipsnames]{color}
\usepackage[pdftex]{graphicx}
\usepackage{tikz}
\usepackage{tabularx}
\usepackage{siunitx}
\usepackage[version=4]{mhchem}
\usetikzlibrary{positioning,arrows.meta,shapes}
\usepackage{pgfplots}
\usepgfplotslibrary{groupplots}
\pgfplotsset{width=6.5cm, compat=1.9}
\usepackage{relsize}

\makeatletter
  \def\my@tag@font{\normalsize}
  \def\maketag@@@#1{\hbox{\m@th\normalfont\my@tag@font#1}}
  \let\amsmath@eqref\eqref
  \renewcommand\eqref[1]{{\let\my@tag@font\relax\amsmath@eqref{#1}}}
\makeatother

\renewcommand{\vec}[1]{\mathbf{#1}}

\newcommand{\norm}[1]{\left\lVert#1\right\rVert}

\newcolumntype{Y}{>{\centering\arraybackslash}X}

\def\tre{^{\mathrm{III}}}

\begin{document}
\title{CERES: An \textit{ab initio} code dedicated to the calculation of the electronic structure and magnetic properties of lanthanide complexes}

\author{Simone Calvello}
\author{Matteo Piccardo}
\author{Shashank Vittal Rao}
\author{Alessandro Soncini}
\email{asoncini@unimelb.edu.au}

\affiliation{School of Chemistry, University of Melbourne, VIC 3010, Australia}

\begin{abstract}
We have developed and implemented a new ab initio code, \textsc{Ceres} (\textsc{C}omputational \textsc{E}mulator of \textsc{R}are \textsc{E}arth \textsc{S}ystems), completely written in C++11, which is dedicated to the efficient calculation of the electronic structure and magnetic properties of the crystal field states arising from the splitting of the ground state spin-orbit multiplet in lanthanide complexes. The new code gains efficiency via an optimised implementation of a direct configurational averaged Hartree-Fock (CAHF) algorithm for the determination of $4f$ quasi-atomic active orbitals common to all multi-electron spin manifolds contributing to the ground spin-orbit multiplet of the lanthanide ion. The new CAHF implementation is based on quasi-Newton convergence acceleration techniques coupled to an efficient library for the direct evaluation of molecular integrals, and problem-specific density matrix guess strategies. After describing the main features of the new code, we compare its efficiency with the current state--of--the--art ab initio strategy to determine crystal field levels and properties, and show that our methodology, as implemented in \textsc{Ceres}, represents a more time-efficient computational strategy for the evaluation of the magnetic properties of lanthanide complexes, also allowing a full representation of non-perturbative spin-orbit coupling effects.
\end{abstract}
\maketitle

\begin{center}
{\bf Graphical Abstract} 
\end{center}

\begin{center}
\includegraphics[width=50mm,height=50mm]{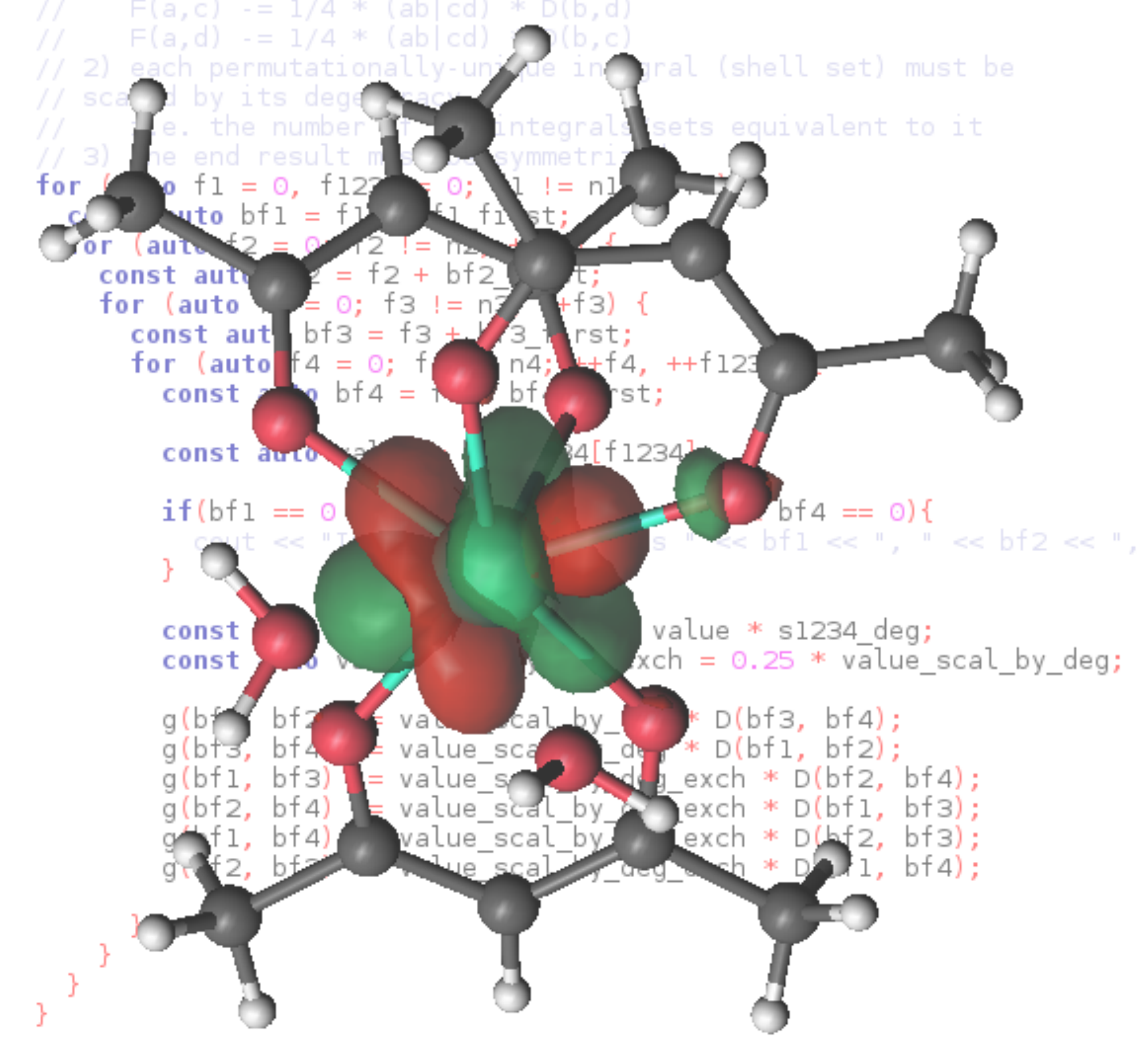}
\\
Lanthanide single molecule magnets are important systems for the development of molecular memories, with ab initio methodologies being an important tool for their characterization. In this work, we present a new software for the calculation of crystal field states for lanthanide single molecule magnets, based on a new method we recently developed, and we compare it with currently available program, showing that our method is more efficient without any significant loss in accuracy.
\end{center}

\section{Introduction}

The magnetic properties of lanthanide single--molecule magnets (SMMs) have made them important targets for a wide range of applications, ranging from magnetic resonance contrast agents~\cite{mrireview} to building blocks for molecular magnetic memories~\cite{lnsmm}. The main reason for their versatility lies in the details of their electronic structure and, in particular, in the almost complete lack of covalent interactions between lanthanide and ligand orbitals for the lowest energy states which, in turn, leads to a quasi--atomic valence space constituted by $\ce{Ln}\tre$ $4f$--like orbitals. Their wave function, therefore, will be dominated by spin--orbit $J$--multiplets weakly split by electrostatic interactions with the ligand crystal field, causing a strong magnetic anisotropy which is fundamental for their multiple applications~\cite{nanomagnets,lnsmm}.

\textit{Ab initio} calculations have proven to be very useful in describing the magnetic properties and identifying magnetostructural correlations for a number of lanthanide complexes displaying SMM behaviour, thus becoming a common feature in experimental studies in the literature. The method of choice for such studies is a combination of Complete Active Space Self--Consistent Field~\cite{casscf1,casscf2} and Restricted Active Space State Interaction via Spin--Orbit coupling~\cite{rassi} (CASSCF/RASSI--SO), as for instance implemented in the quantum chemistry program \textsc{Molcas}~\cite{molcas}. Given a mononuclear $\ce{Ln}\tre$ complex with an open--shell configuration of $4f^n$, out of all multiconfigurational spin states arising from the distribution of $n$ valence electrons in the seven $4f$--like valence orbitals, a subset is chosen and, in the CASSCF step of the calculation, their wave functions are optimized.   Next, the spin-free CASSCF wave functions are used as basis states to diagonalize the spin-orbit coupling (SOC) operator, and obtain the resulting crystal field states (RASSI step). This method was first employed to theoretically explain the presence of a non--magnetic ground state in a $\ce{Dy3}$ triangle~\cite{toroidal}, and has since been used to successfully rationalize the magnetic properties of a range of lanthanide SMMs~\cite{liviu,liviuliviu,doubledecker,cpercot,NatComm2013,vonci}.

Spin--orbit coupling is one of the dominant energy scales in lanthanide complexes so that it would be desirable to use all CASSCF optimized spin states to represent the SOC Hamiltonian matrix, but in current CASSCF implementations this approach can be computationally very demanding. In the RASSI step, furthermore, all CASSCF spin state wave functions have to be converted to a common molecular orbital basis. The latter process can also be computationally demanding, and even in efficient programs like \textsc{Molcas} the maximum number of CASSCF spin states which can be currently handled is around 300~\cite{liviu}, thus not allowing a full representation of the SOC interaction in many experimentally relevant ions like $\ce{Tb}\tre$, $\ce{Dy}\tre$ and $\ce{Ho}\tre$.

In a recent work~\cite{CONFAV} we have discussed the limits of the CASSCF/RASSI--SO approach for lanthanide SMMs and proposed an alternative computational strategy, called CAHF/CASCI--SO (Configurational Average Hartree--Fock/Complete Active Space Configuration Interaction with Spin--Orbit Coupling), which allows the calculation of magnetic properties of $\ce{Ln}\tre$ complexes where all spin states are included. In the preliminary calculations shown there, the new method did not introduce any significant deviation from CASSCF/RASSI--SO crystal field levels and magnetic properties. In this paper, we present a dedicated program for CAHF/CASCI--SO calculations, called \textsc{Ceres}~\cite{ceres_inc}, showcasing its main features and comparing its efficiency with the more established CASSCF/RASSI--SO computational strategy for the crystal field states and magnetic properties of $\ce{Ln}\tre$ complexes, as implemented in the software \textsc{Molcas}. 

\section{CAHF/CASCI--SO theory as implemented in \textsc{Ceres}}

 The current ab initio strategy to determine the crystal field states in $\ce{Ln}\tre$ complexes consists of performing a set of State Averaged CASSCF (SA-CASSCF) calculations~\cite{liviu}. In particular, for each spin quantum number allowed by the occupation of the 4f active space, the SA-CASSCF energy functional used in the orbital optimisation problem is defined as the average energy of all multiconfigurational electronic eigenstates with a given spin. The SOC operator is then represented on the basis of as many SA-CASSCF eigenstates as it is computationally feasible to include, and then diagonalised (RASSI-SO step), yielding the crystal field levels~\cite{liviu}. Note that SA-CASSCF eigenstates with different spins feature different (non-orthogonal) sets of optimised molecular orbitals for different spin manifolds, which can make the calculation inefficient.    Note also that while in a general SA-CASSCF calculation orbital rotations are coupled to the CI coefficients optimization problem, when averaging is carried out over \textit{all} possible multiconfigurational eigenstates within a given active space carrying a specific irreducible representation (irrep) of a given symmetry group (e.g. having a given total spin), the average energy becomes the trace of the electrostatic Hamiltonian matrix within a given irrep of the symmetry group, which, being invariant under a unitary transformation, becomes independent of the eigenstates of the problem. In other words, the orbital optimisation problem becomes exactly decoupled from the CI problem in typical applications of SA-CASSCF to $\ce{Ln}\tre$ complexes, and could in principle be solved as a single CAHF problem.~\cite{McWeenyConfav,McWeeny,CONFAV}  
 
 This simple observation is the starting point for our proposed CAHF/CASCI-SO strategy.  In particular, on account of the fact that strong-spin orbit coupling does not conserve the molecule's spin angular momentum, in the proposed CAHF/CASCI-SO we relax the condition that state-averaging should be carried out within a given spin manifold. Hence, the average energy functional is simply the trace of the CI matrix represented on the basis of all possible Slater determinants, of any $M_S$ quantum number, spanning the chosen active space.  After having thus optimised average 4f orbitals common to all multiconfigurational spin states arising from a given occupation of the  4f active space (CAHF step), we use these orbitals to build a representation of the sum of electrostatic and SOC Hamiltionians on the basis of single Slater determinants (CASCI-SO step), and diagonalise it to obtain the crystal field levels.  The CAHF/CASCI-SO strategy has now been implemented in an efficient C++11 code, \textsc{Ceres}, which is described in some more detail in what follows. 
 
 In the rest of the Section we will provide a formulation of the CAHF procedure using second quantization, which is known to be more amenable to the definition of efficient convergence algorithms.  From this point on, therefore, we will consider a general open--shell lanthanide complex, and we will partition its orbitals into three subspaces: $m_I$ inactive orbitals, always doubly occupied, $m_A$ active orbitals, occupied by $n_A$ electrons, and $m_V$ virtual orbitals, always unoccupied. We will then define average orbital space occupation numbers as $\nu_X = n_X/m_X$ for the subspace $X$. Finally, molecular orbital indices will be partitioned as follows: $i, j, k, l$ for inactive orbitals, $u, v, w, x$ for active orbitals, $a, b, c, d$ for virtual orbitals and $p, q, r, s$ for unspecified orbitals.

The well-known second quantization formulation of Self-Consistent Field theory~\cite{pinkbook} is briefly rehearsed below, in order to expose its specificities when applied to the configurationally averaged problem implemented in the CAHF module of our code \textsc{Ceres}.    An optimal parameterisation of the Born-Oppenheimer electronic energy functional, leading to a formulation of CAHF as an unconstrained energy minimization problem, is achieved by imposing an exponential parameterisation of the (unitary) operator performing orbital rotations in the multi-electron wavefunction. For instance, given the Slater determinant  $\left| \Phi_i \left(\bm{\kappa}\right) \right\rangle$ wavefunction, we have:
\begin{equation}
\label{rotation}
\left| \Phi_i \left(\bm{\kappa}\right) \right\rangle = e^{-\hat{\kappa}} \left| \Phi_i \right\rangle
\end{equation} 
with $\hat{\kappa} = \sum_{p > q} \kappa_{pq} \left(E_{pq} - E_{qp}\right)$, where, according to the usual notation, if $a_{p\sigma}$ is an annihilation operator destroying an electron with spin $\sigma$ in the molecular orbital $\phi_p$, the singlet excitation operator reads $E_{pq}=\sum_{\sigma}a_{p\sigma}^{\dagger}a_{q\sigma}$.   The argument of the Slater determinant wavefunction $\left| \Phi_i \left(\bm{\kappa}\right) \right\rangle$ is the antisymmetric matrix $\bm{\kappa}$,  containing all non--redundant orbital rotation parameters, i.e. rotations between different orbital subspaces (inactive, active and virtual) defined in the CAHF calculation.

On account of strong spin-orbit coupling, and considering that in our target systems the active space spans the 4f atomic angular momentum shell well-shielded from covalency effects by 5s and 5p electrons, especially for the smaller lanthanide ions of interest to molecular magnetism due to their large ground state angular momentum, we assume it is physically reasonable to fully relax the spin-symmetry of the system, and define the CAHF energy functional as the average energy of all the Slater determinants $\left| \Phi_i \left(\bm{\kappa}\right) \right\rangle$, $i = 1,\dots, N_{\mathrm{SD}}$, spanning the chosen 4f active space (i.e. for single-ion lanthanide complexes, $N_{\mathrm{SD}}=\binom{14}{n}$, where $n$ is the number of 4f electrons in the chosen lanthanide ion), irrespective of their $M_S$ quantum number. This leads to the following expression for the energy:
\begin{align}
\overline{E}\left(\bm{\kappa}\right) = & N_{\mathrm{SD}}^{-1} \sum_{i}^{N_{\mathrm{SD}}} \left\langle \Phi_i \left(\bm{\kappa}\right) \right|  H_{el}  \left|  \Phi_i \left(\bm{\kappa}\right) \right\rangle \nonumber\\
                                 = & \sum_{pq} h_{pq} \overline{D}_{pq} + \frac{1}{2} \sum_{pqrs} g_{pqrs} \overline{d}_{pqrs} + E_{\mathrm{nuc}} \label{energyconfavexplicit}
\end{align}
where $H_{el}$ is the molecular electrostatic Hamiltonian, $h_{pq}$ and $g_{pqrs}$ are the mono-- and bi--electronic molecular integrals in the molecular orbital basis expressed in Mulliken notation, and the mono-- ($\overline{D}_{pq}$) and bi--electronic ($\overline{d}_{pqrs} $) \textit{average} density matrices in the molecular orbital basis, defined as:
\begin{align}
\overline{D}_{pq} = & N_{\mathrm{SD}}^{-1} \sum_i^{N_{\mathrm{SD}}} \sum_{\sigma} \left\langle \Phi_i \left|  a^\dag_{p\sigma} a_{q\sigma} \right| \Phi_i \right\rangle \nonumber \\
\overline{d}_{pqrs} = & N_{\mathrm{SD}}^{-1} \sum_i^{N_{\mathrm{SD}}} \sum_{\sigma,\sigma'} \left\langle\Phi_i \right|  a^\dag_{p\sigma} a^\dag_{r\sigma'} a_{s\sigma'} a_{q\sigma} \left| \Phi_i \right\rangle,
\label{density}
\end{align}
are here explicitly calculated as:
\begin{align}
\label{1ed_CONFAV}
\overline{D}_{pq} = & 
\begin{cases}
\nu_I \delta_{pq} & p \text{ inactive} \\
\nu_A \delta_{pq} & p \text{ active} \\
0 & p \text{ virtual}
\end{cases}
\end{align}
\begin{small}
\begin{align}
\label{2ed_CONFAV}
\overline{d}_{pqrs} = & 
\begin{cases}
\nu_I \left(2 \delta_{pq} \delta_{rs} - \delta_{ps} \delta_{rq}\right) & p \text{, } r \text{ inactive} \\
\nu_I \nu_A \left(\delta_{pq} \delta_{rs} - \frac{1}{2} \delta_{ps} \delta_{rq}\right) & p \text{ inactive, } r \text{ active} \\
 & \text{or } \\
 & p \text{ active, } r \text{ inactive} \\
\nu_A \lambda \left(\delta_{pq} \delta_{rs} - \frac{1}{2} \delta_{ps} \delta_{rq}\right) & p \text{, } r \text{ active} \\
0 & \text{ otherwise}
\end{cases}
\end{align}
\end{small}
with $\lambda = \frac{n_A - 1}{2m_A - 1}$ being the average active space electron--electron repulsion weight.

Substitution of Eqs.~(\ref{1ed_CONFAV}--\ref{2ed_CONFAV}) into the average energy expression Eq.~(\ref{energyconfavexplicit}) gives the CAHF energy expression~\cite{McWeenyConfav,CONFAV}:
\begin{align}
\label{CONFAV_energy}
\overline{E} & = \nu_I \sum_i \left[h_{ii} + \sum_j \left(g_{iijj} - \frac{1}{2} g_{ijji}\right)\right] \nonumber \\
& + \nu_I \nu_A \sum_{i,u} \left(g_{iiuu} - \frac{1}{2} g_{iuui}\right) \\
& + \nu_A \sum_u \left[h_{uu} + \frac{\lambda}{2} \sum_v \left(g_{uuvv} - \frac{1}{2} g_{uvvu}\right)\right] + E_{nuc} \nonumber
\end{align}

 Minimization of Eq.~(\ref{CONFAV_energy}) with respect to the matrix of rotation parameters leads to the usual linear system of equations whose solution defines the family of convergence algorithms known as second--order methods:
\begin{equation}
\label{second_order}
\overline{\mathbf{E}}^{(2)} \bm{\kappa} = - \overline{\mathbf{E}}^{(1)}
\end{equation}
 where the antisymmetric matrix $\bm{\kappa}$ collecting the orbital rotations is here arranged in a column vector, and now $\overline{\mathbf{E}}^{(2)}_{pq,rs} = \left.\frac{\partial^2 \overline{E}}{\partial \kappa_{pq} \partial \kappa_{rs}}\right|_{\bm{\kappa}=0}$ is the configurationally averaged molecular Hessian and $\overline{\mathbf{E}}^{(1)}_{pq} = \left.\frac{\partial \overline{E}}{\partial \kappa_{pq}}\right|_{\bm{\kappa}=0}$ is the configurationally averaged molecular gradient, which can be shown to read:
\begin{subequations}
\begin{align}
\label{gradient_CAHF}
\overline{E}^{(1)}_{iu} = & 2 \left(\nu_I \overline{F}^{(1)}_{iu} - \nu_A \overline{F}^{(2)}_{ui}\right) \\
\overline{E}^{(1)}_{ia} = & 4 \overline{F}^{(1)}_{ia} \\
\overline{E}^{(1)}_{ua} = & 2 \nu_A \overline{F}^{(2)}_{ua}
\end{align}
\end{subequations}
with
\begin{equation}
\begin{aligned}
\label{genfomaexp_inac}
\overline{F}^{(1)}_{pq} = & h_{qp} + \nu_I \sum_i \left(g_{qpii} - \frac{1}{2} g_{qiip}\right) \\
& + \nu_A \sum_u \left(g_{qpuu} - \frac{1}{2} g_{quup}\right)
\end{aligned}
\end{equation}
and
\begin{equation}
\begin{aligned}
\label{genfomaexp_ac}
\overline{F}^{(2)}_{pq} = & h_{qp} + \nu_I \sum_i \left(g_{qpii} - \frac{1}{2} g_{qiip}\right) \\
& + \nu_A \lambda \sum_u \left(g_{qpuu} - \frac{1}{2} g_{quup}\right)
\end{aligned}
\end{equation}
The expression for the configurationally averaged molecular Hessian $\overline{\mathbf{E}}^{(2)}$ is reported in the Supplementary Informations, along with a more thorough derivation of the gradient $\overline{\mathbf{E}}^{(1)}$. Since the computation of Hessian matrix elements is extremely time--consuming, a direct solution of Eq.~(\ref{second_order}), either by inverting the Hessian or by solving iteratively the linear system of equations, is rarely employed. A significant speed--up is provided by the well--known Quasi--Newton method~\cite{restricted}, based on the secant condition, in which an approximate Hessian is used to solve Eq.~(\ref{second_order}), and at every iteration the knowledge provided by the gradients is used to improve such approximation, with the most used update scheme being the Broyden-–Fletcher-–Goldfarb-–Shanno (BFGS) algorithm~\cite{broyden,fletcher,goldfarb,shanno}:
\begin{align}
\label{bfgs}
\mathbf{H}_{k+1} = & \mathbf{H}_k + \frac{\left(\mathbf{s}_k^\dag \mathbf{y}_k + \mathbf{y}_K^\dag \mathbf{H}_k \mathbf{y}_k\right)\left(\mathbf{s}_k \mathbf{s}_k^\dag\right)}{\left(\mathbf{s}_k^\dag \mathbf{y}_k\right)^2} \nonumber \\
& - \frac{\mathbf{H}_k \mathbf{y}_k \mathbf{s}_k^\dag + \mathbf{s}_k \mathbf{y}_k^\dag \mathbf{H}_k}{\mathbf{s}_k^\dag \mathbf{y}_k}
\end{align}
with $\mathbf{H}_k$ being the approximated inverted Hessian, $\mathbf{y}_k = \mathbf{\mathbf{E}}^{(1)}_{k + 1} - \mathbf{\mathbf{E}}^{(1)}_k$ and $\mathbf{s}_k = \mathbf{\kappa}^{k + 1} - \mathbf{\kappa}^k$ at iteration $k$.

Finally, $H_{el}+H_{\mathrm{SOC}}$, where $H_{\mathrm{SOC}}$ is an appropriate representation of the SOC Hamiltonian  (\textit{vide infra}), is represented on the basis of the $N_{\mathrm{SD}}$ Slater determinants (SD's) built with the CAHF--optimized molecular orbitals.   $H_{el}+H_{\mathrm{SOC}}$ is then diagonalized to yield the crystal field energies and multiconfigurational wavefunctions.  Note that, while $H_{el}$ is block diagonal in the full SD basis  \textit{i.e.} it has matrix elements only between SD's with like $M_S$ quantum numbers, $H_{\mathrm{SOC}}$ features first-rank irreducible tensor spin operators coupling a given $M_S$ SD subspace, to the two SD subspaces characterized by $M_S\pm1$ quantum numbers. Given the basis of SD's which, for a given $M_S$ quantum number, can be naturally represented as strings of $\alpha$ and $\beta$ electrons partially occupying the active space molecular orbitals, the CI routine is naturally based on the $\sigma$--algorithm developed by Olsen~\cite{olsenci,pinkbook}, with the addition of spin non-conserving single excitations to represent the one-electron SOC Hamiltonian:
\begin{equation} 
        H_{\mathrm{SOC}} = \sum_i \vec{l}(i)\cdot\vec{s}(i), 
\end{equation} 
where the summation is over all electrons. In second quantization this gives
\begin{multline} 
        H_{\mathrm{SOC}}  =\frac{1}{2} \sum_{pq} l^z_{pq} (a^\dagger_{p\alpha}a_{q\alpha} - a^\dagger_{p\beta}a_{q\beta}) \\
        + l^+_{pq} a^\dagger_{p\beta}a_{q\alpha} +  l^-_{pq} a^\dagger_{p\alpha}a_{q\beta}
\end{multline} 
where $l^z_{pq}$ and $l^{\pm}_{pq} = l^x_{pq}\pm i l^y_{pq}$ are the integrals of the (bare or mean--field) one-electron SOC Hamiltonian in the active molecular orbital basis representation. Generalization of the algorithm to include two-electron SOC contributions is underway~\cite{barespinorbit}.

\section{\textsc{Ceres} Structure}

In this Section, we present \textsc{Ceres}~\cite{ceres_inc}, a quantum chemistry code specifically designed for efficient CAHF/CASCI-SO calculations of crystal field states and magnetic properties of lanthanide complexes. After discussing the \textsc{Ceres} main features, we test its performances and efficiency with respect to the so far adopted SA-CASSCF/RASSI-SO strategy.

The program \textsc{Ceres} is written in C++11, taking advantage of object-oriented programming. Implemented by making use of the open-source C++ \textsc{Boost} libraries~{\cite{boost}}, a \textsc{Python} front-end makes \textsc{Ceres} user-friendly, and automates many common tasks such as parsing the input data, building jobs as a sequence of desired methods, the re--execution of a sequence of calculations on a set of molecules, or a quick and direct control of the output data.   All matrix operations are performed using the fast and efficient C++ \textsc{Eigen} template library for linear algebra~{\cite{eigenweb}}. A particular attention has been dedicated to the parallelization of the code where appropriate, by the use of openMP application programming interface (API) specification for parallel programming.


\tikzset{
     startstop/.style={ellipse, draw, fill=white, text width=6em, text centered, minimum height=3em},
     mine/.style={rectangle, draw, fill=white, text width=6em, text centered, minimum height=3em},
     decision/.style={diamond, draw, fill=white, text width=5em, text centered, minimum height=2em},
     arrow/.style={draw, thick, color=black!75, -latex}
     }

\begin{figure*}[tbp]
\includegraphics{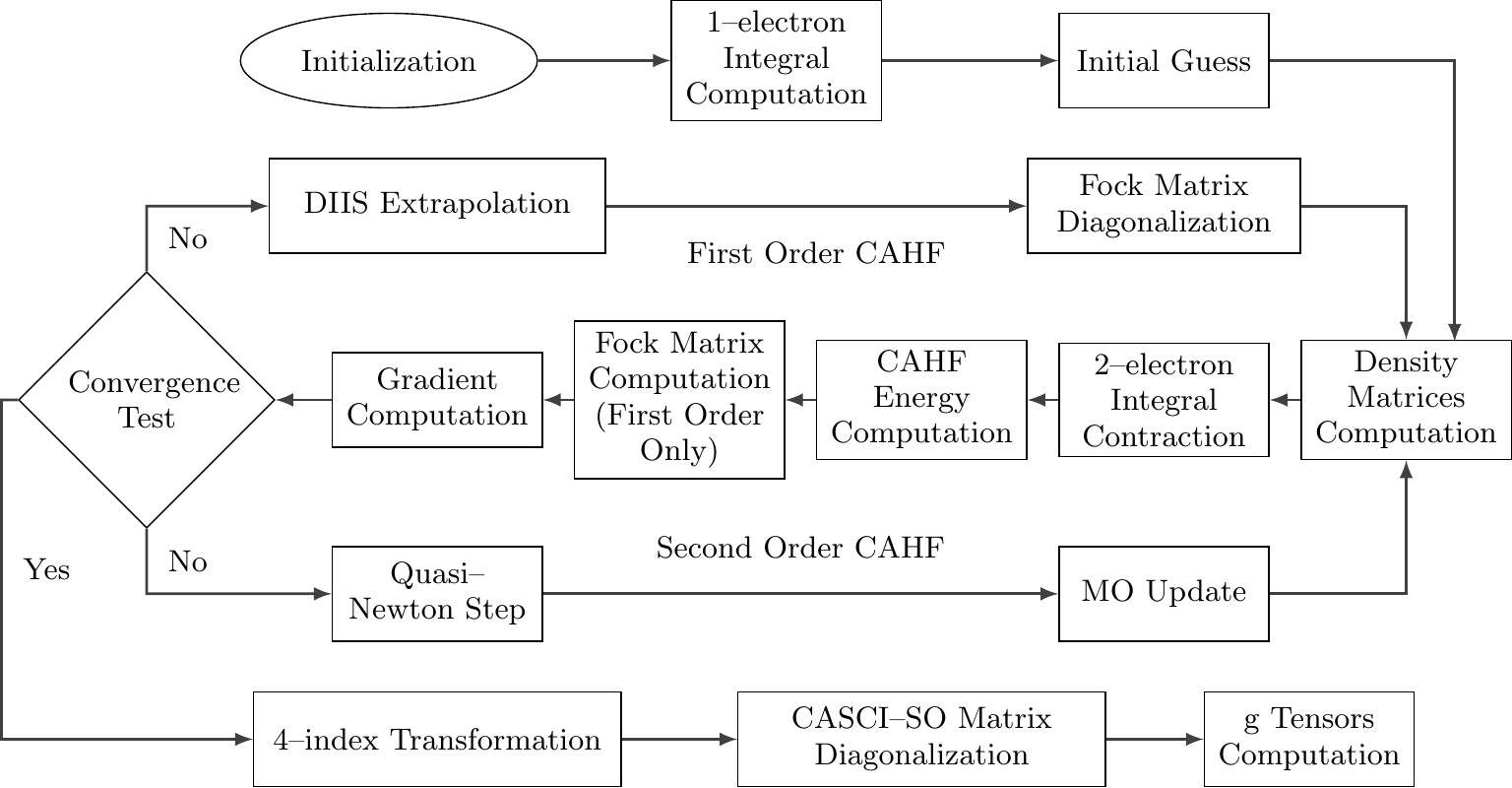}
\captionof{figure}{Flow chart of \textsc{Ceres}.}
\label{ceresflowchart}
\end{figure*}
 
The structure of the implemented CAHF/CASCI-SO algorithm is sketched in Fig.~\ref{ceresflowchart}.
The CAHF function consists of a direct SCF module, where the integral calculation at each iteration always occurs on the fly, using the \textsc{Libint} library for high-performance Gaussian integrals computation~{\cite{libint}}, which is based on the Obara-Saika recursion scheme.  A version of the code which is currently being debugged is including the more general and efficient \textsc{Libcint} library for Gaussian basis functions~\cite{libcint}, based on the Rys-polynomials scheme, to overcome some \textsc{Libint} limitations in handling multiple contractions, and to implement all the relativistic and spin-orbit integrals, which are of crucial importance to accurately describe lanthanide complexes.   All direct CAHF calculations presented here have been performed with the \textsc{Libint} library, as this implementation has been extensively debugged.  At this stage, the relativistic integrals entering both the scalar Douglas-Kroll-Hess (DKH) corrections to the one-electron Hamiltonian~{\cite{DKH}} and the one-electron SOC Hamiltonian have been calculated by the use of an \textit{ad hoc} modified version of the \textsc{Libint} C++ library. For the purpose of comparing our numerical results with those obtained via the CASSCF/RASSI--SO strategy as implemented in \textsc{Molcas}, we decided to read in our CASCI module the SOC AMFI integrals calculated with \textsc{Molcas} in the calculations presented here, postponing the comparison between different approximations of the SOC interactions to a separate forthcoming study.~\cite{barespinorbit,amfienc}


\textsc{Ceres} features several algorithms for the estimation of the initial orbital guess feeding the CAHF iterative calculation, some of which are well known, namely the diagonalization of the mono--electronic Hamiltonian and the projection on a larger basis set of the density matrices obtained from a CAHF calculation on a smaller basis set. Since previous studies indicated that simple charge models are quite effective in predicting the properties of lanthanide complexes~\cite{NatComm2013}, however, we chose to implement a generalized form of the Sum of Atomic Densities (SOAD) guess~\cite{vanlenthe} specifically designed for calculations of $\ce{Ln}\tre$ SMM magnetic properties, in which we assume that the initial atomic orbitals are orthogonal and localized. The resulting density matrices will thus be diagonal, with their non--zero elements corresponding to the occupation number of the atomic orbital whose angular momentum is $l$, computed as $\nu_{n,l} = \frac{n}{4l + 2}$. Chemical knowledge of the target molecule can be translated into input partial charges to be distributed on selected atoms or ligands. Density matrices generated with this procedure are, then, converted into the non--orthogonal basis using an inverse L\"{o}wdin transformation:
\begin{equation}
\label{invlow}
\textbf{R}^X_{non-orth} = \textbf{S}^{-\frac{1}{2}} \textbf{R}^X_{orth} \textbf{S}^{-\frac{1}{2}}
\end{equation}
where $\textbf{S}$ is the overlap matrix in the atomic orbital basis.
\textsc{Ceres} also allows to create such diagonal density matrices in a smaller basis set and project them in the final basis set via a mixed--basis bi--electronic integral contraction.

The direct SCF algorithm we have implemented~\cite{directscf} features a preliminary screening of the list of overlap distributions for a given basis set, achieved  by computing the overlap between basis set shells, and by avoiding computation of all integrals between such shells if, provided they do not belong to the same atom, the Euclidean norm of their overlap is lower than a chosen threshold.  Secondly, at each iteration a Cauchy--Schwarz screening procedure is employed~\cite{cauchy}, with a dynamical screening threshold defined so as to improve the precision of the calculation closer to convergence:
\begin{equation}
\label{threshold}
\delta = min \left( min \left( \frac{1}{S_{cond}}, 10^{-9} \right), max \left( \frac{\norm{\bar{\mathbf{E}}^{(1)}}_\infty}{10^{-7}}, \epsilon \right) \right)
\end{equation}
where $S_{cond}$ is the condition number of the overlap matrix, $\epsilon$ is the machine precision, and $\norm{\mathbf{x}}_\infty = max\left(\left|x_1\right|, \left|x_2\right|, \cdots\right)$.  Finally, \textsc{Ceres} uses an incremental Fock build algorithm~\cite{incremental}, which becomes particularly efficient when coupled with Cauchy--Schwarz screening because the difference between consecutive iterations density matrices is, in general, smaller when closer to convergence. Incremental Fock build is started when sufficiently close to convergence and, since it introduces errors in the energy, it is reset every eight iterations by performing a non--incremental iteration. Precision of the final result is granted by turning off incremental Fock build when all convergence criterions are smaller than an order of magnitude higher than the convergence thresholds. The effective Fock Hamiltonian, which is diagonalized in every first--order iteration~\cite{McWeeny, CONFAV}, includes level shifters $\lambda_1$ and $\lambda_2$~\cite{levelshift} so as to separate the energies of orbital subspaces and improve the efficiency of the calculation, which are defined as:
\begin{equation}
\label{shifters}
\bar{\mathbf{F}}\left(\lambda_1, \lambda_2\right) = \bar{\mathbf{F}} + \lambda_1 \mathbf{R}_2 + \left(\lambda_1 + \lambda_2\right) \mathbf{R}_3
\end{equation}
where $\bar{\mathbf{F}}$ is the effective CAHF Hamiltonian and $\mathbf{R}_2$ and $\mathbf{R}_3$ are, respectively, the active and virtual density matrices in the atomic orbital basis. Convergence criterions are the same used in \textsc{Molcas}~\cite{molcas}, namely $\Delta \bar{E}$ and $\norm{\bar{E}^{(1)}}_\infty$. 

\textsc{Ceres} features a mixed first--second order convergence algorithm, using DIIS~\cite{diis,diis2} with the molecular gradient as the error choice~\cite{ioncar} in the first part. Once sufficiently close to convergence, using $\norm{\bar{E}^{(1)}}_\infty$ and $\norm{\Delta \bar{R}}_\infty$ as criterions, with $\bar{R}$ being the average density matrix in the atomic orbital basis, the second--order convergence algorithm is used. \textsc{Ceres} implements both an iterative solution of Eq.~(\ref{second_order}) via the Jacobi method and a Limited--Memory BFGS Quasi--Newton algorithm~\cite{lbfgs}, whose main advantage with respect to the BFGS algorithm is the possibility to compute the updated Hessian without storing it in memory, thus improving the scalability of the method. Although most QN algorithms use identity matrix as initial Hessian approximation, we have chosen to use the one--electron component of the diagonal elements of the exact molecular Hessian so as to improve the efficiency of QN and avoid time--consuming line--search methods~\cite{linesearch}.

Once CAHF convergence has been achieved, the optimized LCAO coefficient matrix is used to set up the CASCI--SO matrix. In the current version of \textsc{Ceres}, spin--orbit contributions are either based on the LIBCINT library, either using the Breit-Pauli Hamiltonian, soon to be interfaced with the Cholesky decomposition of the bi--electronic spin--orbit integrals~\cite{barespinorbit}, or by employing the well--known AMFI (Atomic Mean--Field Integrals) approximation~\cite{amfi} in various formulations, including the possibility to read them from files produced by \textsc{Molcas}, as done here for ease of comparison. The optimized LCAO coefficient matrix is, then, used to transform electron repulsion bi--electronic integrals into the molecular orbital active subspace via two successive two--index semi--transformations~\cite{fourindex}.  Cholesky decomposition of the electron repulsion integrals is also going to be soon used in this step, so that the integrals on the molecular basis are going to be re-composed as a sum of Cholesky vectors transformed on the molecular basis. However, results presented here do not yet make use of this facility. The electron-repulsion and spin-orbit coupling integrals transformed on the molecular basis are finally used to build the matrix for the CASCI--SO problem, the implementation of which is based on the $\sigma$--algorithm developed by Olsen~\cite{olsenci}.

\section{Analysis of \textsc{Ceres} Performance}

\begin{figure*}[tpb]
\begin{minipage}{\textwidth}
\begin{minipage}{0.59\textwidth}
\begin{tabular}{|c|c|c|c|}
\hline
             &    \multirow{2}{*}{Molecule}    &         SA--CASSCF/         &        CAHF/       \\
             &        &         RASSI--SO         &        CASCI--SO       \\
\hline
\multirow{4}{*}{\textbf{1}} &   $\ce{[Tb(acac)3(H2O)2]}$   &          06:55         &        02:00      \\
             &   $\ce{[Dy(acac)3(H2O)2]}$   &          08:57         &        02:55      \\
             &   $\ce{[Ho(acac)3(H2O)2]}$   &          09:59         &        03:37      \\
             &   $\ce{[Er(acac)3(H2O)2]}$   &          03:06         &        03:02      \\
\hline
\multirow{4}{*}{\textbf{2}}  &   $\ce{[Tb(acac)3(dppz)]}$   &          26:45         &        06:12      \\
             &   $\ce{[Dy(acac)3(dppz)]}$   &          28:01         &        08:20      \\
             &   $\ce{[Ho(acac)3(dppz)]}$   &          34:02         &        09:21      \\
             &   $\ce{[Er(acac)3(dppz)]}$   &          13:42         &        07:36      \\
\hline
\multirow{4}{*}{\textbf{3}}  &   $\ce{[Tb(acac)3(dpq)]}$   &          22:15         &        05:17      \\ 
             &   $\ce{[Dy(acac)3(dpq)]}$   &          21:33         &        09:12      \\
             &   $\ce{[Ho(acac)3(dpq)]}$   &          25:17         &        08:08      \\
             &   $\ce{[Er(acac)3(dpq)]}$   &          09:51         &        06:05      \\
\hline
\multirow{4}{*}{\textbf{4}}  &   $\ce{[Tb(acac)3(phen)]}$   &          20:29         &        05:11      \\
             &   $\ce{[Dy(acac)3(phen)]}$   &          18:47         &        07:15      \\ 
             &   $\ce{[Ho(acac)3(phen)]}$   &          21:52         &        07:39      \\ 
             &   $\ce{[Er(acac)3(phen)]}$   &          06:57         &        06:59      \\
\hline
\multirow{4}{*}{\textbf{5}}  &   $\ce{[Tb(hfac)3(dme)]}$   &          20:46         &        04:29      \\
             &   $\ce{[Dy(hfac)3(dme)]}$   &          23:20         &        06:08      \\
             &   $\ce{[Ho(hfac)3(dme)]}$   &          27:46         &        07:29      \\
             &   $\ce{[Er(hfac)3(dme)]}$   &          11:35         &        04:53      \\
\hline
\multirow{4}{*}{\textbf{6}}  &   $\ce{[Tb(paaH^{*})2(NO3)2(MeOH)]+}$   &          18:48         &        04:10      \\
             &   $\ce{[Dy(paaH^{*})2(NO3)2(MeOH)]+}$   &          17:18         &        05:46      \\
             &   $\ce{[Ho(paaH^{*})2(NO3)2(MeOH)]+}$   &          19:05         &        07:12      \\
             &   $\ce{[Er(paaH^{*})2(NO3)2(MeOH)]+}$   &          09:01         &        04:22      \\
\hline
\multirow{4}{*}{\textbf{7}}  &   $\ce{[Tb(tfpb)3(dppz)]}$   &          99:35         &        13:44      \\
             &   $\ce{[Dy(tfpb)3(dppz)]}$   &          88:26         &        18:43      \\
             &   $\ce{[Ho(tfpb)3(dppz)]}$   &          97:53         &        13:39      \\
             &   $\ce{[Er(tfpb)3(dppz)]}$   &          42:03         &        12:15      \\
\hline
\multirow{4}{*}{\textbf{8}}  &   $\ce{[Tb(tta)3(bipy)]}$   &          50:41         &        10:09      \\
             &   $\ce{[Dy(tta)3(bipy)]}$   &          45:04         &        09:34      \\
             &   $\ce{[Ho(tta)3(bipy)]}$   &          40:06         &        08:46      \\
             &   $\ce{[Er(tta)3(bipy)]}$   &          20:07         &        08:30      \\
\hline
\multirow{4}{*}{\textbf{9}}  &   $\ce{[Tb(tta)3(phen)]}$   &          52:23         &        09:09      \\
             &   $\ce{[Dy(tta)3(phen)]}$   &          48:54         &        13:27      \\
             &   $\ce{[Ho(tta)3(phen)]}$   &          39:04         &        08:55      \\
             &   $\ce{[Er(tta)3(phen)]}$   &          26:20         &        09:54      \\
\hline
\multirow{4}{*}{\textbf{10}} &   $\ce{[Tb(tta)3(pinene-bipy)]}$   &          71:41         &        11:49      \\
             &   $\ce{[Dy(tta)3(pinene-bipy)]}$   &          74:26         &        14:19      \\
             &   $\ce{[Ho(tta)3(pinene-bipy)]}$   &          55:27         &        11:58      \\
             &   $\ce{[Er(tta)3(pinene-bipy)]}$   &          35:25         &        10:57      \\
\hline
\end{tabular}
\captionof{table}{Timings, expressed as $hh$:$mm$, for the crystal field level calculations of the chosen complexes by SA--CASSCF/RASSI-SO and CAHF/CASCI--SO.}
\label{timings}
\end{minipage}
\begin{minipage}{0.39\textwidth}
\includegraphics{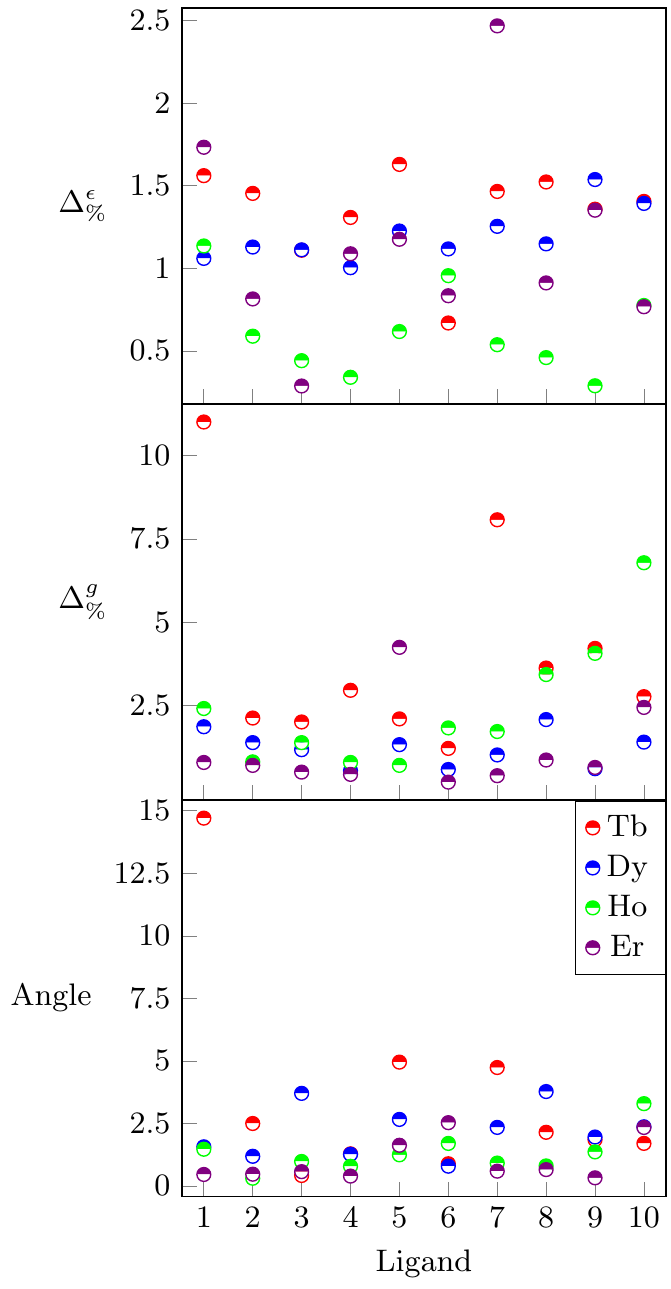}
\captionof{figure}{Plots for the accuracy study of the SA--CASSCF/RASSI-SO, and CAHF/CASCI--SO methods. The three plots showcase the accuracy of, respectively, crystal field levels (top), g--tensors (middle) and anisotropy axes (bottom). Ligand number on the $x$ axis is extracted from Table~\ref{timings}.}
\label{accuracy}
\end{minipage}
\end{minipage}
\end{figure*}

Although CAHF/CASCI--SO and SA--CASSCF/RASSI--SO are two conceptually different methods that cannot be directly compared in terms of the speed of each single step in the process, both are expected to lead to results of comparable accuracy, and an efficient implementation of our proposed CAHF/CASCI--SO is expected to be competitive with the SA--CASSCF/RASSI--SO approach for the calculation of the electronic structure and magnetic properties of the crystal field levels arising from the ground spin--orbit term of any $\ce{Ln}\tre$ complex. In order to provide evidence of this, and to test the efficiency of our implementation of the CAHF/CASCI--SO method, we have performed single point calculations on a set of ten complexes of $\ce{Tb}\tre$, $\ce{Dy}\tre$, $\ce{Ho}\tre$ and $\ce{Er}\tre$ with \textsc{Ceres}, comparing the performance with the CASSCF/RASSI--SO strategy as implemented in \textsc{Molcas} 8.0~\cite{molcas}. Experimental geometries were available for all $\ce{Dy}$ compounds~\cite{NatComm2013} and, whenever possible, experimental geometries have been used for other ions~\cite{Lnacac,dyacacphen,dyacacdppz,Lnhfac,chilton2013,dytfnb3dppz,Dytta3bipy}. All calculations for each ion have been performed on separate NeCTAR research cloud virtual machines~\cite{nectar}, each with $16$ $\si{\giga}$B RAM and $2.3$ $\si{\giga\hertz}$ Intel CPUs. 

The active space is made of the seven $\ce{Ln}\tre$ $4f$ orbitals occupied by, respectively, 8 electrons in $\ce{Tb}$, 9 electrons in $\ce{Dy}$, 10 electrons in $\ce{Ho}$ and 11 electrons in $\ce{Er}$. Given the inability of LIBINT to efficiently handle general contraction basis sets, we have chosen segmented basis sets, namely SARC2--QZVP--DKH for $\ce{Ln}\tre$~\cite{sarc2}, Ahlrichs--PVDZ for coordinating atoms and Ahlrichs--VDZ for all other atoms~\cite{ahlrichs}. Throughout the paper, the following abbreviations will be in use to shorten the names of the molecules: $\ce{acac}$ = acetylacetonate, $\ce{dppz}$ = dipyridophenazine, $\ce{dpq}$ = dipyridoquinoxaline, $\ce{phen}$ = 1,10-phenanthroline, $\ce{hfac}$ = hexafluoroacetylacetone, $\ce{dme}$ = dimethoxyethane, $\ce{paaH^*}$ = N-(2-Pyridyl)acetoacetamide, $\ce{tfpb}$ = 4,4,4-trifluoro-1-phenyl-1,3-butandionate, $\ce{tta}$ = tetradecylthioacetate, $\ce{bipy}$ = 2,20-bipyridine, $\ce{pinene-bipy}$ = 4,5-pinene bipyridine.

\textsc{Molcas} calculations are performed using High Cholesky option, with a cut--off threshold of $10^{-8}$. 
We have performed one SA--RASSCF calculation for each possible total spin quantum number $S$ by averaging over all possible spin states except for $\ce{Tb}$ triplets, for which the number of states has been reduced so as not to exceed virtual machine memory. The total number of SA--CASSCF calculations performed for each lanthanide complex is, respectively, four for $\ce{Tb}$, three for $\ce{Dy}$ and $\ce{Ho}$ and two for $\ce{Er}$.
In the RASSI step, we have selected all states from some of the lowest Russell--Saunders terms for each $S$ while still maintaining a total number of states less than 300~\cite{liviu}. An overview of the number of states used in each step is presented in the Supplementary Informations. 

\textsc{Ceres} calculations include a CAHF calculation on all spin states with level shifters for active and virtual space of $0.4$ and convergence criterions set to the same values of \textsc{Molcas} as $\Delta E \leq 10^{-8}$ and $\norm{E^{(1)}}_\infty \leq 10^{-4}$. 

\begin{figure*}[tbp]
\includegraphics{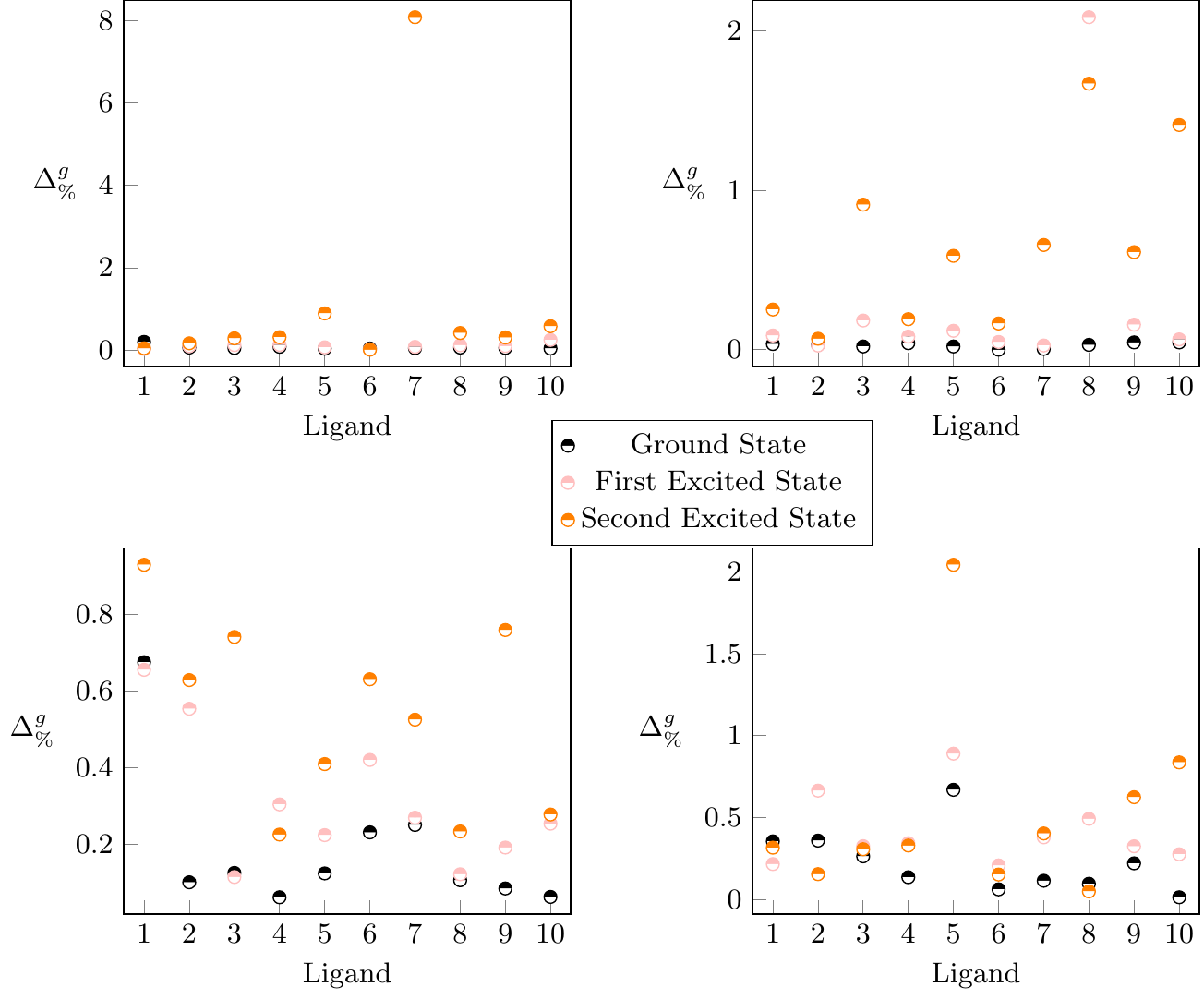}
\captionof{figure}{Plots of the relative error for the g--tensor computation for the ground state and the first two excited Kramers doublets, for odd--electron systems, or Ising doublets, for even--electron systems, between the SA--CASSCF/RASSI-SO, and CAHF/CASCI--SO methods. The four plots represent, respectively, the g--tensors for $\ce{Tb}$ (top left), $\ce{Dy}$ (top right), $\ce{Ho}$ (bottom left) and $\ce{Er}$ (bottom right). Ligand number on the $x$ axis is extracted from Table~\ref{timings}.}
\label{accuracyg}
\end{figure*}

\begin{table*}[tpb]
\begin{tabular}{|c|c||c|c||c|c|}
\hline
 & Molecule & SA--CASSCF & RASSI--SO & CAHF & CASCI--SO \\
\hline
\multirow{4}{*}{\textbf{1}} &   $\ce{[Tb(acac)3(H2O)2]}$   & 04:50 & 02:05 & 01:30 & 00:30 \\
             &   $\ce{[Dy(acac)3(H2O)2]}$   & 05:16 & 03:41 & 02:20 & 00:35 \\
             &   $\ce{[Ho(acac)3(H2O)2]}$   & 08:40 & 01:19 & 02:53 & 00:44 \\
             &   $\ce{[Er(acac)3(H2O)2]}$   & 02:33 & 00:33 & 02:33 & 00:29 \\
\hline
\multirow{4}{*}{\textbf{2}}  &   $\ce{[Tb(acac)3(dppz)]}$   & 20:19 & 06:26 & 05:02 & 01:10 \\
             &   $\ce{[Dy(acac)3(dppz)]}$   & 20:39 & 07:22 & 06:44 & 01:36 \\
             &   $\ce{[Ho(acac)3(dppz)]}$   & 30:14 & 03:48 & 07:50 & 01:31 \\
             &   $\ce{[Er(acac)3(dppz)]}$   & 11:29 & 02:13 & 06:25 & 01:11 \\
\hline
\multirow{4}{*}{\textbf{3}}  &   $\ce{[Tb(acac)3(dpq)]}$   & 17:57 & 04:18 & 04:17 & 01:00 \\
             &   $\ce{[Dy(acac)3(dpq)]}$   & 15:40 & 05:53 & 07:29 & 01:43 \\
             &   $\ce{[Ho(acac)3(dpq)]}$   & 22:43 & 02:34 & 06:50 & 01:18 \\
             &   $\ce{[Er(acac)3(dpq)]}$   & 07:50 & 02:01 & 04:52 & 01:13 \\
\hline
\multirow{4}{*}{\textbf{4}}  &   $\ce{[Tb(acac)3(phen)]}$   & 15:52 & 04:37 & 04:16 & 00:55 \\
             &   $\ce{[Dy(acac)3(phen)]}$   & 12:59 & 05:48 & 06:04 & 01:11 \\
             &   $\ce{[Ho(acac)3(phen)]}$   & 19:59 & 01:53 & 06:11 & 01:28 \\
             &   $\ce{[Er(acac)3(phen)]}$   & 05:59 & 00:58 & 05:55 & 01:04 \\
\hline
\multirow{4}{*}{\textbf{5}}  &   $\ce{[Tb(hfac)3(dme)]}$   & 15:32 & 05:14 & 03:32 & 00:57 \\
             &   $\ce{[Dy(hfac)3(dme)]}$   & 16:45 & 06:35 & 04:39 & 01:29 \\
             &   $\ce{[Ho(hfac)3(dme)]}$   & 24:18 & 03:28 & 06:08 & 01:21 \\
             &   $\ce{[Er(hfac)3(dme)]}$   & 09:25 & 02:10 & 03:59 & 00:54 \\
\hline
\multirow{4}{*}{\textbf{6}}  &   $\ce{[Tb(paaH^{*})2(NO3)2(MeOH)]+}$   & 14:38 & 04:10 & 03:25 & 00:45 \\
             &   $\ce{[Dy(paaH^{*})2(NO3)2(MeOH)]+}$   & 11:53 & 05:25 & 04:41 & 01:05 \\
             &   $\ce{[Ho(paaH^{*})2(NO3)2(MeOH)]+}$   & 17:19 & 01:46 & 05:51 & 01:21 \\
             &   $\ce{[Er(paaH^{*})2(NO3)2(MeOH)]+}$   & 07:19 & 01:42 & 03:32 & 00:50 \\
\hline
\multirow{4}{*}{\textbf{7}}  &   $\ce{[Tb(tfpb)3(dppz)]}$   & 87:20 & 12:15 & 11:07 & 02:37 \\
             &   $\ce{[Dy(tfpb)3(dppz)]}$   & 73:33 & 14:53 & 14:46 & 03:57 \\
             &   $\ce{[Ho(tfpb)3(dppz)]}$   & 91:44 & 06:09 & 11:04 & 02:35 \\
             &   $\ce{[Er(tfpb)3(dppz)]}$   & 37:33 & 04:30 & 09:42 & 02:33 \\
\hline
\multirow{4}{*}{\textbf{8}}  &   $\ce{[Tb(tta)3(bipy)]}$   & 42:27 & 08:14 & 08:14 & 01:55 \\
             &   $\ce{[Dy(tta)3(bipy)]}$   &  34:42 & 10:22 & 07:41 & 01:53 \\
             &   $\ce{[Ho(tta)3(bipy)]}$   &  35:26 & 04:40 & 06:52 & 01:54 \\
             &   $\ce{[Er(tta)3(bipy)]}$   & 17:21 & 02:46 & 06:41 & 01:49 \\
\hline
\multirow{4}{*}{\textbf{9}}  &   $\ce{[Tb(tta)3(phen)]}$   & 42:28 & 09:55 & 07:12 & 01:57 \\
             &   $\ce{[Dy(tta)3(phen)]}$   & 37:12 & 11:42 & 10:42 & 02:45 \\
             &   $\ce{[Ho(tta)3(phen)]}$   & 34:13 & 04:51 & 07:03 & 01:52 \\
             &   $\ce{[Er(tta)3(phen)]}$   & 22:48 & 03:32 & 08:03 & 01:51 \\
\hline
\multirow{4}{*}{\textbf{10}} &   $\ce{[Tb(tta)3(pinene-bipy)]}$   & 60:04 & 11:37 & 09:43 & 02:06 \\
             &   $\ce{[Dy(tta)3(pinene-bipy)]}$   & 59:51 & 14:35 & 11:40 & 02:39 \\
             &   $\ce{[Ho(tta)3(pinene-bipy)]}$   & 49:45 & 05:42 & 09:27 & 02:31 \\
             &   $\ce{[Er(tta)3(pinene-bipy)]}$   & 31:45 & 03:40 & 08:37 & 02:20 \\
\hline
\end{tabular}
\caption{Timings, expressed as $hh$:$mm$, for the main steps of the calculations required to compute the crystal field levels of the chosen complexes using the SA--CASSCF/RASSI-SO, and CAHF/CASCI--SO methods.}
\label{timingsdiss}
\end{table*}

In order to discuss both the efficiency and accuracy of the CAHF/CASCI--SO method we will analyze, respectively, the timings required for the magnetic properties calculation and the values of crystal field energies, g--tensors and orientation of the main anisotropy axes for the ground state spin--orbit multiplet. For odd--electron systems, g--tensors and anisotropy axes will be computed between degenerate states of each Kramers doublet, while for even--electron systems they will be computed between two non--degenerate, consecutive states. 
We have reported the computational timings for the overall calculations with the SA--CASSCF/RASSI-SO and CAHF/CASCI--SO methods in Table~\ref{timings}, with a breakdown into the timings for each phase of all calculations in Table~\ref{timingsdiss}. 
All values of crystal field energies and g--tensors for the ground state spin--orbit term of each molecule have been included in the Supplementary Informations. In order to analyze and discuss the accuracy of our method with respect to SA--CASSCF/RASSI-SO, then, we have computed the relative difference between crystal field energies and between the largest components of the g--tensors as, respectively, 
\[
\Delta^{\epsilon}_{\%} = \frac{\left|\epsilon_{\textsc{Molcas}} - \epsilon_{\textsc{Ceres}}\right|}{max\left(\epsilon_{\textsc{Molcas}}, \epsilon_{\textsc{Ceres}}\right)} \cdot 100 
\]
\[
 \Delta^g_{\%} = \frac{\left|g_{\textsc{Molcas}} - g_{\textsc{Ceres}}\right|}{max\left(g_{\textsc{Molcas}}, g_{\textsc{Ceres}}\right)} \cdot 100 
 \]
We have also computed the angle between the main anisotropy axes computed with the two methods. 
We have, then, collected data subsets into several plots to highlight the most important conclusions from this study. We have included the highest errors for crystal field levels, g--tensors and anisotropy axes in Fig.~\ref{accuracy}, the relative errors for the g--tensor for the ground crystal field state and the first two excited states for all the 40 molecules in Fig.~\ref{accuracyg} and the relative errors for the crystal field energy of the second excited state, which corresponds to a state in the first excited Kramers doublet for odd--electron systems and to the lower energy state in the first excited pseudo--Kramers doublet for even--electron systems, in Fig.~\ref{accuracyc}. 

The main conclusion which can be drawn from the analysis of Table~\ref{timings} is that our implementation of the CAHF/CASCI--SO algorithm is indeed more efficient than SA--CASSCF/RASSI--SO. Out of the 40 molecules considered, in fact, the latter displays similar timings only for $\ce{[Er(acac)3(H2O)2]}$ and $\ce{[Er(acac)3(phen)]}$. 

Despite introducing a significative speed--up, the CAHF/CASCI--SO method does not introduce significant losses in accuracy. Fig.~\ref{accuracy} does, in fact, show remarkable agreement for the crystal field levels between the two methods, with an average deviation of $\Delta^{\epsilon}_{\%} = 1.5{\%}$ and the highest error being $2.47\%$ for $\ce{[Er(tfpb)3(dppz)]}$. Analysis of g--tensors also displays a good agreement between methods, with the lowest errors arising for odd--electron systems. Even--electron systems show slightly higher errors, with three notable outliers in $\ce{[Tb(acac)3(H2O)2]}$, $\ce{[Tb(tfpb)3(dppz)]}$ and $\ce{[Ho(tta)3(pinene-bipy)]}$. Anisotropy axes are also reproduced very well by the CAHF/CASCI--SO method, with the maximum error being lower than $5\%$ for all molecules but one. 

It has to be noted that the most relevant deviations between values usually occur for high energy crystal field states, which is confirmed by the plots in Fig.~\ref{accuracyg}. When only the lowest three Kramers or Ising doublets are considered, deviations between methods are greatly reduced, with relative errors usually lower than $2\%$ but for $\ce{[Tb(tfpb)3(dppz)]}$. Since these states are the most significant for explaining several magnetic phenomena, we can thus safely claim that the CAHF/CASCI--SO method will be accurate in describing them. 

\begin{center}
\begin{minipage}{\columnwidth}
\includegraphics{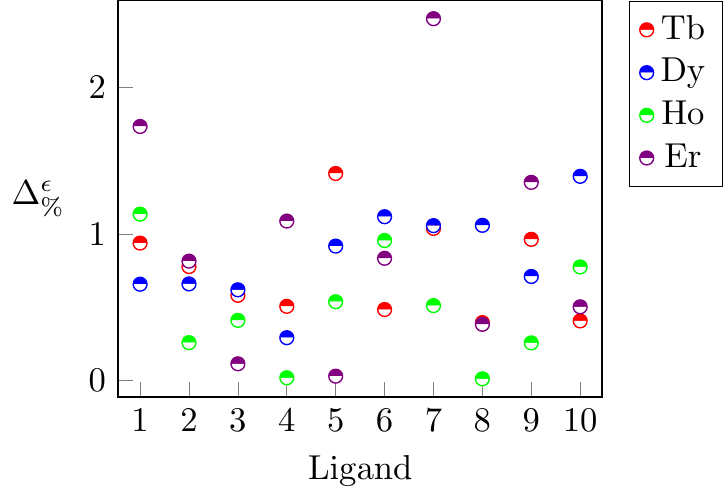}
\captionof{figure}{Plots of the relative error of the energy of the first excited Kramers doublet, for odd--electron systems, or of the lower energy state of the first excited Ising doublet, for even--electron systems, between the SA--CASSCF/RASSI-SO, and CAHF/CASCI--SO methods. Ligand number on the $x$ axis is extracted from Table~\ref{timings}.}
\label{accuracyc}
\end{minipage}
\end{center}

Finally, we note that the energy of the lowest excited crystal field state is also well reproduced, as shown in Fig.~\ref{accuracyc}, with deviations averaging $1\%$. Overall, these results indicate that the CAHF/CASCI--SO method, as implemented in \textsc{Ceres}, provides an efficient alternative to the more established CASSCF/RASSI--SO method with the approximations introduced not leading to big errors.




\begin{center}
{\bf Supplementary Informations} 
\end{center}

Supplementary Informations are available for this Article. They include a Table with the number of CASSCF and RASSI roots employed in all calculations, a full list of all abbreviations used with respect to the SMM structures, details on the derivation of molecular gradient and Hessian, geometries used for all calculations in this paper and crystal field states and g--tensor values for the ground state spin--orbit multiplet of all compounds studied above.

\begin{center}
{\bf Acknowledgements} 
\end{center}
 
All authors gratefully acknowledge support from the Australian Research Council, Discovery Project grant ID: DP150103254.  S.C. and S.V.R. thankfully acknowledge support from an Australian Government Research Training Program Scholarship.  This research was supported by use of the Nectar Research Cloud, a collaborative Australian research platform supported by the National Collaborative Research Infrastructure Strategy (NCRIS).
 

%
\end{document}